\documentclass{elsart}
\usepackage{graphics}
\usepackage{graphicx}
\usepackage{epsfig}
\usepackage{amssymb}
\usepackage{amsmath}
\begin{document}

\begin{frontmatter}

\title{The measurement of the $\mathbf{pp{\to}K^+n\Sigma^+}$
reaction near threshold}

\author[PNPI,IKP]{Yu.~Valdau}\ead{y.valdau@fz-juelich.de},
\author[UCL]{C.~Wilkin\corauthref{cor1}}
\ead{cw@hep.ucl.ac.uk}
\corauth[cor1]{Corresponding author.}

\address[PNPI]{High Energy Physics Department, Petersburg Nuclear
  Physics Institute, RU-188350 Gatchina, Russia}
\address[IKP]{Institut f\"ur Kernphysik, Forschungszentrum J\"ulich,
  D-52425 J\"ulich, Germany}
\address[UCL]{Physics and Astronomy Department, UCL, London WC1E 6BT, UK}

\begin{abstract}
It is shown that a recent extraction of the total cross section for
$pp\to K^+n\Sigma^+$ from inclusive $K^+$ production data is in
conflict with experimental data on the exclusive $pp\to K^+p\Lambda$
reaction. The result may be interpreted as an upper bound which is
not inconsistent with the much lower values that already exist in the
literature.
\end{abstract}

\end{frontmatter}

Two recent measurements of the total cross sections for the
$pp{\to}K^+n\Sigma^+$ reaction close to threshold at
COSY-ANKE~\cite{VAL2007,VAL2010} and COSY-HIRES~\cite{BUD2010} have
given conflicting results. It is the purpose of this short note to
suggest reasons for this discrepancy.

Following an initial study at the COSY-ANKE spectrometer at an excess
energy of $\varepsilon = 129$~MeV~\cite{VAL2007}, data on
$pp{\to}K^+n\Sigma^+$ were obtained at four energies closer to
threshold~\cite{VAL2010}. This was achieved by measuring in parallel
coincidence spectra from $K^+$-proton and $K^+\pi^+$ pairs, the
latter being especially convincing because at low energies they can
only arise from $\Sigma^+$ production. The results could be checked
by studying the ratio of inclusive $K^+$ production in $pp$
collisions just above to just below the $\Sigma^+$ threshold. The
three methods gave consistent answers and showed production cross
sections that were slightly smaller than those of
$pp{\to}K^+p\Sigma^{\,0}$, with $R(\Sigma^+/\Sigma^0) \approx
0.7\pm0.1$. The energy variation found was consistent with a
phase-space dependence, with no evidence for a strong $\Sigma^+n$
final state interaction (FSI).

In an alternative approach, the COSY-HIRES
collaboration~\cite{BUD2010} measured with high resolution the
forward inclusive production of $K^+$ mesons in proton-proton
collisions at a single beam momentum of $p=2.870$~GeV/$c$,
corresponding to an excess energy for $\Sigma^+$ production of
$\varepsilon=102.6$~MeV. Following a procedure developed
earlier~\cite{SIB2007}, they claimed that, after the subtraction of
an assumed excitation function for the $\Lambda$ channel, their
results showed that $R(\Sigma^+/\Sigma^0) \approx 5\pm 1$. However,
the extraction of a $pp{\to}K^+n\Sigma^+$ cross section from such
data is not straightforward because, due to the mass difference, the
available phase space is much larger for $\Lambda$ production.
Furthermore, even at the same excess energy the cross section for
$\Sigma^0$ production is typically over an order of magnitude smaller
than that of the $\Lambda$~\cite{SEW1999}

The above difficulties are compounded by the strong coupling between
the $\Sigma N$ and $\Lambda p$ channels, which can lead to a cusp in
the $\Lambda p$ spectrum at the $\Sigma N$ threshold. A spectacular
example of this is seen in the $K^-d\to \pi^-\Lambda p$
reaction~\cite{TAN1969}. However, because such effects are sensitive
to the interference between direct $\Lambda p$ production and that
proceeding via the formation of a virtual $\Sigma N$ pair, the form
of the $\Lambda p$ spectrum can vary greatly from one reaction to
another. This is illustrated by the comparison of pion
photoproduction near the $\eta p$ threshold, with very contrasting
structure being seen in the $\gamma p\to \pi^+n$ and $\gamma p\to
\pi^0p$ cross sections~\cite{ALT1983}. It is the purpose of this note
to point out that new exclusive data on $pp{\to}K^+p\Lambda$,
obtained at the COSY Time-of-Flight spectrometer
(COSY-TOF)~\cite{ABD2010}, show an anomalous behaviour near the $\Sigma N$
threshold. The particular form of this undermines the assumptions
inherent in the COSY-HIRES analysis~\cite{BUD2010} and suggests that
their value for the $pp{\to}K^+n\Sigma^+$ total cross section should
be interpreted only as an upper limit.

\begin{figure}[htb]
\begin{center}
\resizebox{12cm}{!}{\includegraphics[scale=1]{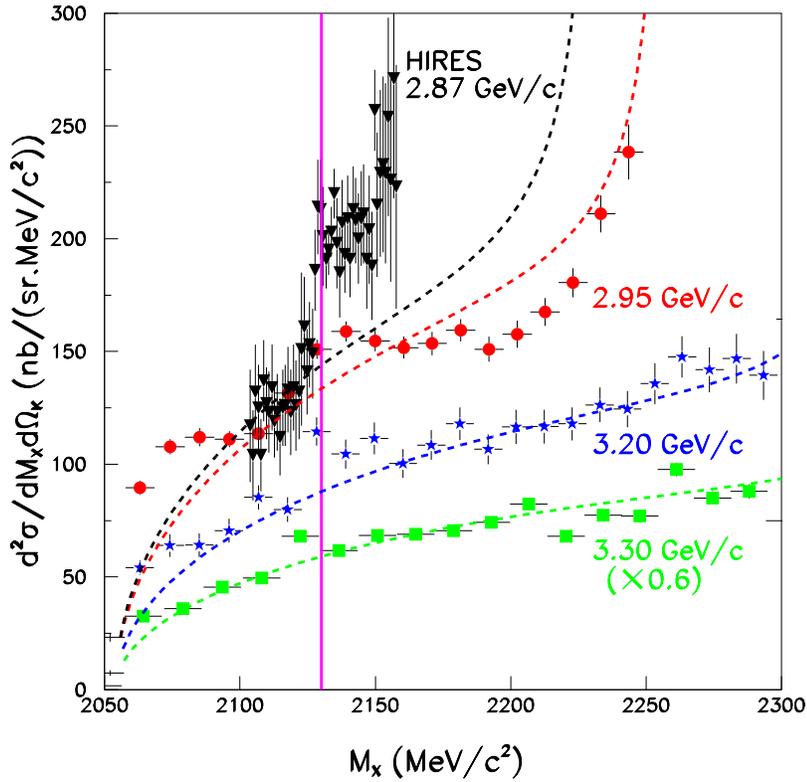}}
\caption{Forward inclusive cross section for $pp\to K^+X$ as a
function of the missing mass $M_X$ in the reaction. The HIRES data at
a beam momentum of 2.87~GeV/$c$ (inverted triangles)~\cite{BUD2010}
were directly measured. The COSY-TOF data at 2.95, 3.20, and
3.30~GeV/$c$ were derived from exclusive $pp\to K^+p\Lambda$
measurements~\cite{ABD2010} by assuming that the distribution in the
$K^+$ c.m.\ angle is isotropic. For clarity of presentation, the
3.30~GeV/$c$ data have been scaled by a factor of 0.6. The vertical
line indicates the position of the average $\Sigma^+n/\Sigma^0p$
threshold. Also shown are arbitrarily normalised three-body
phase-space distributions corresponding to the $pp\to K^+p\Lambda$
reaction.
 \label{money-plot} }
\end{center}
\end{figure}

Inclusive $K^+$ production in proton-proton collisions was studied
with the SPES4 spectrometer at SATURNE at fixed non-zero laboratory
angles~\cite{SIE1994}. Clear evidence was there obtained for a strong
rise in the $pp\to K^+X$ differential cross section in the vicinity
of the $\Sigma N$ threshold. These studies were taken up later at the
Big Karl spectrometer, which was situated on an external beam line of
the COSY accelerator. Working only in the forward direction, the
HIRES collaboration~\cite{BUD2010,BUD2010a} obtained a better
missing-mass $M_X$ resolution than that of the SPES4 experiment. They
could also determine quite accurately the absolute value of $M_X$ and
hence its relation to the $\Sigma N$ threshold. The resulting forward
differential cross sections are presented in Fig.~\ref{money-plot} as
a function of $M_X$.

The strong rise in the cross section in Fig.~\ref{money-plot}, which
starts well below the $\Sigma N$ threshold, must be associated with
$\Lambda$ production that is driven by coupled channel effects. The
SPES4 data could be described in a phenomenological meson-exchange
model by Laget~\cite{LAG1991}, though the resultant structure in the
threshold region is quite complex and his assignment of the
production strength into different channels above the $\Sigma N$
threshold is very model-dependent. There is also the possibility of
fine structure, with a double cusp arising from the differences
between the $\Sigma^0p$ and $\Sigma^+n$ thresholds~\cite{DAL1993}.

Although these HIRES data are very detailed, their interpretation is
contentious. It is assumed, without independent evidence, that the
structure of the $pp{\to}K^+p\Lambda$ cusp contribution is a simple
Breit-Wigner form with a full width of $\approx 6$~MeV/$c^2$. It is
further hypothesised that, away from the near-threshold region, the
differential cross section for $pp{\to}K^+p\Lambda$ is essentially
the same to the left and right of the $\Sigma N$ threshold, in marked
contrast even to the phase-space behaviour, shown by the dashed lines in
Fig.~\ref{money-plot}. The inevitable
consequence of these assumptions is that the excess of cross section
to the right of the near-threshold region must be due to $\Sigma$
production. However, since it is known that $\Sigma^0$ production
through $pp{\to}K^+p\Sigma^{\,0}$ is relatively modest, this would
have to be compensated by a large value for the
$pp{\to}K^+n\Sigma^{+}$ cross section.

Fortunately, exclusive COSY-TOF data on the $pp\to K^+ p \Lambda$
reaction allow us to test the hypothesis
underpinning this analysis. In Refs.~\cite{ABD2010, ABD2006} the measurement
and analysis of Dalitz plots and $p\Lambda$ invariant masses at 2.75,
2.85, 2.95, 3.2 and 3.3~GeV/$c$ are presented. The most prominent
features of the Dalitz plots are the band associated
with the excitation of the $N^*(1650)$ and an enhancement in
the vicinity of the $\Sigma N$ threshold~\cite{ABD2010, ABD2006}.
Though the data at 2.75 and 2.85~GeV/$c$~\cite{ABD2006} have relatively
large fluctuations, both show structure at the $\Sigma N$ threshold.
However, the authors note that ``the precision of the data is not
sufficient to draw any conclusion on this''. On the other hand
data published by the COSY-TOF collaboration earlier this year,
with their most refined results on the exclusive cross section for
$pp\to K^+p\Lambda$ at three beam momenta, viz.\ 2.95, 3.20,
and 3.30~GeV/$c$~\cite{ABD2010}, confirm presence of the structure
in the vicinity of the $\Sigma$ production threshold. The most
pronounced effect is observed in the highest statistics data set
collected at 2.95~GeV/$c$~\cite{ABD2010}.

Because the TOF spectrometer has a very large acceptance, the data
obtained with it are spread over the whole of the three-body phase
space and it is not meaningful to try to extract results with a small
$K^+$ angular cut. However, after summing over all $\Lambda p$
invariant masses $M_{\Lambda p}$, it is found that globally the $K^+$
angular distribution is rather isotropic, whereas those in the proton
and $\Lambda$ show a peaking towards the forward/backward
directions~\cite{ABD2010a}. The HIRES analysis~\cite{BUD2010} goes a
little further than this in that it is there assumed that the cross
section is independent of the $K^+$ c.m.\ angle for all values of
$M_X$. This reasonable assumption allows them to derive total cross
sections from the values of the forward $K^+$ differential cross
sections. By the same token, it allows us to estimate forward $K^+$
differential cross sections from the angle-integrated exclusive
COSY-TOF data~\cite{ABD2010} and the results of doing this are shown
in Fig.~\ref{money-plot}~\cite{SIB2009}.

Also shown in the figure are arbitrarily normalised $pp\to
K^+p\Lambda$ phase-space distributions corresponding to the HIRES and
the three COSY-TOF momenta. For kinematic reasons these distributions
increase strongly with $M_X$ and the effect of this in the $\Sigma N$
threshold region becomes more marked as the beam momentum is reduced
and the maximum missing mass gets closer. The rapidly varying phase
space changes the peak seen in the 2.95~GeV/$c$ COSY-TOF integrated
cross section~\cite{ABD2010} into a smeared step function in the
forward laboratory cross section of Fig.~\ref{money-plot}.

The statistics of the COSY-TOF data and the inherent resolution lead
to a much wider binning than that used for the HIRES results but a
rise in the cross section around the $\Sigma N$ threshold is clear.
In addition to the kinematic effect discussed above, this
might be influenced by the overlap of the $N^*(1650)$ with the
$\Sigma N$ threshold region then representing a larger fraction of
the Dalitz plot.

The HIRES data at 2.87~GeV/$c$ show a lot of similarities to the
COSY-TOF results at 2.95~GeV/$c$ and some of the residual differences
might be due to the 80~MeV/$c$ offset in beam momentum and the
normalisation uncertainties, as well as the $K^+$ isotropy
assumption. In both cases the sharp rise in the cross section starts
at about 10~MeV/$c^2$ below the $\Sigma N$ threshold and well to the
right of the threshold the cross sections stay high. To quantify
this, let us consider the change in the cross sections in the stable
regions from the left to the right of the threshold. The increase for
the HIRES inclusive data is $\approx 60\%$ whereas for the COSY-TOF
exclusive results at 2.95~GeV/$c$ the change is closer to $\approx
40\%$. However, it should be noted that a 40\% rise in the forward
$pp\to K^+p\Lambda$ cross section from 2110 to 2160~MeV/$c^2$ is not
very different from that which is predicted at 2.95~GeV/$c$ on the
basis of phase space, as shown in Fig.~\ref{money-plot}. The channel
coupling seems merely to sharpen up the steady phase-space rise in
this representation.

Although the HIRES 2.87~GeV/$c$ and the COSY-TOF 2.95~GeV/$c$ data
themselves look somewhat similar, the interpretation offered is
starkly different. The HIRES group assumed that the differential
cross section for $pp{\to}K^+p\Lambda$ is essentially the same on the left and
right of the $\Sigma N$ threshold. This is in flat contradiction to
the COSY-TOF \underline{exclusive} $pp{\to}K^+p\Lambda$ data which
show, at a close beam momentum, that $\Lambda$ production does
not drop to the pre-existing level once the $\Sigma N$ threshold has
been passed. As a consequence, a large fraction of what the HIRES
collaboration has ascribed to $\Sigma$ production might be
$\Lambda$ production, whose shape does not resemble phase space.

Comparison of exclusive COSY-TOF data measured at
2.85~GeV/$c$~\cite{ABD2006} with simulations based on the HIRES 
ansatz show moderate agreement of data and calculations outside the 
regions of the $p\Lambda$ final-state interaction and $\Sigma$ 
production threshold~\cite{BUD2010a}. However, this approach 
fails badly for the high statistics data set at 2.95~GeV/$c$ 
collected later by the same experimental group~\cite{ABD2010}.

Without a clear understanding of the coupled-channel effects or
reliable experimental data on $pp{\to}K^+p\Lambda$ at the HIRES beam
momentum, it is not possible to deduce a model-independent cross
section for $\Sigma$ production from inclusive $K^+$ data, even if
one accepts that this is isotropic in the c.m.\ system. Any 
uncertainty in the $\Lambda$ production cross section is reflected
directly as a systematic error in the $\Sigma$ cross section. The HIRES
analysis~\cite{BUD2010} assumes that, away from a narrow region
around the $\Sigma N$ threshold, the laboratory cross section for
$\Lambda$ production in Fig.~\ref{money-plot} is constant over their
range of missing masses. However, the COSY-TOF data~\cite{ABD2010}
show that $\Lambda$ production generally increases over this region
and that the rise through the $\Sigma N$ threshold is even sharper
than phase space. As a consequence the HIRES analysis probably
provides an upper bound on the cross sections for $\Sigma N$
production, i.e.\ $R(\Sigma^+/\Sigma^0) < 5\pm 1$. Although this is
sufficient to rule out the early results of the COSY-11
group~\cite{ROZ2006} by a large factor, it is of limited value
because the already published COSY-ANKE paper presented the much
lower figure of $R(\Sigma^+/\Sigma^0)=0.7\pm 0.1$ by using three
different experimental techniques at several energies~\cite{VAL2010}.

To show the sensitivity of the HIRES analysis to the assumed
threshold behaviour, if the excess to the right of the $\Sigma N$
threshold in Fig.~\ref{money-plot} were reduced from 60\% to 20\% to
take account of the $pp\to K^+p\Lambda$ jump found in the COSY-TOF
data~\cite{ABD2010}, one would conclude that the total cross sections
for $pp{\to}K^+n\Sigma^+$ and $pp{\to}K^+p\Sigma^0$ would be broadly
similar. This would not be inconsistent with the COSY-ANKE
findings~\cite{VAL2010}.\\

Vigorous interactions within the ANKE collaboration, especially with
C.~Hanhart, A.~Kacharava, V.~P.~Koptev and H.~Str\"{o}her, prompted
this work. Discussions with A.~Sibirtsev regarding the transformation
between the COSY-TOF and COSY-HIRES kinematics were very helpful. We
are grateful to W.~Eyrich and W.~Schroeder for providing us with the
numerical values of the COSY-TOF data of Ref.~\cite{ABD2010}.
Valuable discussions with H.~Clement, S.~Schadmand and J.~Ritman
are gratefully acknowledged. This work was supported by the JCHP FFE. 
%
%

%
%
\end{document}